\documentclass[twocolumn]{aastex63}
\usepackage{gensymb}
\usepackage{hyperref}

\accepted{June 17, 2021}
\submitjournal{PSJ}

\shorttitle{Oxygen Distribution on Ganymede}
\shortauthors{Trumbo et al.}

\graphicspath{{./}{figures/}}

\begin{document}

\title{The Geographic Distribution of Dense-phase O$_2$ on Ganymede}

\correspondingauthor{Samantha K. Trumbo}
\email{strumbo@caltech.edu}

\author{Samantha K. Trumbo}
\affiliation{Division of Geological and Planetary Sciences, California Institute of Technology, Pasadena, CA 91125, USA}

\author{Michael E. Brown}
\affiliation{Division of Geological and Planetary Sciences, California Institute of Technology, Pasadena, CA 91125, USA}

\author{Danica Adams}
\affiliation{Division of Geological and Planetary Sciences, California Institute of Technology, Pasadena, CA 91125, USA}



\begin{abstract}

Ground-based spectroscopy of Ganymede's surface has revealed the surprising presence of dense-phase molecular oxygen (O$_2$) via weak absorptions at visible wavelengths. To date, the state and stability of this O$_2$ at the temperatures and pressures of Ganymede's surface are not understood. Its spatial distribution in relation to albedo, expected temperatures, particle irradiation patterns, or composition may provide clues to these unknowns. We present spatially resolved observations of Ganymede's surface O$_2$ obtained with the Hubble Space Telescope and construct the first comprehensive map of its geography. In agreement with the limited spatially resolved data published previously, our map suggests that the condensed O$_2$ is concentrated at the low- to mid-latitudes of the trailing hemisphere, a distribution that may reflect influences of Ganymede's intrinsic magnetic field on the bombardment of its surface by Jovian magnetospheric particles. Overlapping regions from different observations within our dataset also show evidence for moderate temporal variability in the surface O$_2$, but we are unable to distinguish between potential causes with the available data.
\end{abstract}

\keywords{Galilean satellites (627), Ganymede (2188), Planetary surfaces (2113), Surface composition (2115)}


\section{Introduction} \label{sec:intro}

Molecular oxygen (O$_2$) is a major product of the water-ice radiolytic cycles on Europa, Ganymede, and Callisto. The bombardment of their surface ice by energetic particles within Jupiter’s magnetosphere dissociates the water molecules, resulting in the subsequent formation of O$_2$ \citep[e.g.][]{JohnsonQuickenden1997,JohnsonEtAl2003, CooperEtAl2003}, which dominates the tenuous atmospheres of all three bodies \citep{HallEtAl1998,CunninghamEtAl2015,Hall1995, McGrath2004}. Curiously, condensed O$_2$ has also been detected on the surfaces of all three satellites via characteristic absorptions at 5773 and 6275 \text{\AA} in ground-based disk-integrated spectra \citep{Spencer1995,SpencerCalvin2002}. The detected features stem from the simultaneous excitation of two adjacent O$_2$ molecules, resulting in ${}^{1}\Delta_{g}$ + ${}^{1}\Delta_{g}$ \textleftarrow ${}^{3}\Sigma_{g}^{-}$ + ${}^{3}\Sigma_{g}^{-}$ electronic transitions \citep{Landau1962}. Thus, they are exclusively  observed in dense-phase oxygen (solid, liquid, or high-pressure gas), which poses a puzzle for their presence in spectra of the icy Galilean satellites. At the temperature of the triple point of O$_2$ (54 K), which is well below the daytime temperatures on all three satellites \citep{Spencer1987}, the O$_2$ vapor pressure is 1.46 mbar \citep{Haynes2018}, many orders of magnitude greater than the satellites' pico- to nano-bar surface pressures \citep{McGrath2004}. Thus, solid O$_2$ is not expected to be stable on their surfaces, leading to the suggestion that the surface O$_2$ may be trapped in bubbles or crystal defects within the ice \citep{JohnsonJesser1997,CalvinEtAl1996, Johnson1999comment}.

To date, the state of the O$_2$ is not understood, with suggestions ranging from the aforementioned “microatmospheres” of O$_2$ bubbles \citep{JohnsonJesser1997}, to O$_2$-bearing mixed clathrates \citep{Hand2006}, and even to solid O$_2$ within localized cold traps on the surface or in the near subsurface \citep{VidalEtAl1997, Baragiola1998}. Furthermore, the mechanisms controlling the production and stability of O$_2$ are similarly uncertain. Clues may lie in the spatial distribution of O$_2$ on each satellite, which could reveal enlightening relationships with albedo, composition, particle bombardment, or temperature. 

Observing such geographic patterns is most feasible for Ganymede, as its O$_2$ bands reach $\sim$10 times the strength of those on Europa and Callisto \citep{Spencer1995, SpencerCalvin2002}. Rotationally resolved ground-based spectra of Ganymede suggest that Ganymede's O$_2$ follows a roughly sinusoidal pattern in longitude, with a peak near the center of the trailing hemisphere \citep{Spencer1995}. Limited spatially resolved observations of the trailing hemisphere obtained with the Hubble Space Telescope (HST) Faint Object Spectrograph (FOS) and Wide Field Planetary Camera 2 (WFPC2) have further suggested that the O$_2$ is preferentially concentrated at low- to mid-latitudes \citep{CalvinSpencer1997}. Here, we use spatially resolved HST observations to produce the first comprehensive map of Ganymede's O$_2$ across nearly the entire surface, allowing for widespread comparison with other geographic patterns.

\section{Observations} \label{sec:observations}
We present HST observations of Ganymede's surface O$_2$ made with the Space Telescope Imaging Spectrograph (STIS) across four HST visits. These observations include archival data from three visits in 1999 and one newly executed in 2020. Table \ref{table:obs} lists the corresponding dates, times, and geometries of each visit. In order to obtain complete spatially resolved coverage, the 52$^{\prime\prime}$ x 0.1$^{\prime\prime}$ slit was stepped across Ganymede's disk in 0.1$^{\prime\prime}$ or 0.075$^{\prime\prime}$ increments for the 1999 and 2020 observations, respectively. Each slit-scan pattern was executed in the G750L first-order spectroscopy mode (R$\sim$500) with exposure times of 61 seconds (for the 1999 visits) or 41 seconds (for the 2020 visit) at each slit position. For our analysis, we used the flux- and wavelength-calibrated data provided by HST after standard reduction with the STIS calibration pipeline (calstis). We extracted single spectra by taking individual rows from the two-dimensional spectral images, corresponding to the 0.05$^{\prime\prime}$ pixel-scale (comparable to the the 0.06$^{\prime\prime}$ or $\sim$200-km diffraction-limited resolution at 5773 \text{\AA}). We then calculated the corresponding latitude/longitude coordinates of each extracted pixel using the aperture geometry information from the HST FITS headers and the phase and angular size of Ganymede from JPL Horizons.

One of the slit positions from the Sep. 14 1999 visit (centered on the northeastern limb of Ganymede) appeared to have encountered problems during the data collection, as it lacked signal comparable to neighboring slits or the opposite limb. Thus, we leave it out of our analysis below. Additionally, the new Oct. 6 2020 observation was slightly mispointed due to a failure during coarse guide star acquisition. However, fine guide star acquisition succeeded, and the slit scan still captured most of Ganymede's surface. We obtained a correction to offset the mispointing by constructing a brightness map using the coordinates obtained with the FITS headers, identifying the prominent impact crater Tros, and calculating the pointing error (which is perpendicular to the slit) using Tros' known coordinates. We applied this correction when obtaining the final coordinates for the 2020 visit.

\begin{table}
\begin{center}
\caption{Table of Observations\label{table:obs}}
\begin{tabular}{ccccc}
\hline\\[-4mm] \hline
Date&Time&Central&Central&Angular\\
(UT)&(Start/End)&Lon.&Lat.&Diameter\\ \hline
1999 Aug 28 & 22:50/00:21 &  300 & 3.17 & 1.66$^{\prime\prime}$\\
1999 Sep 23 & 09:41/11:22 &  193 & 3.22 & 1.78$^{\prime\prime}$\\
1999 Sep 14 & 08:21/10:08 &  97 & 3.21 & 1.74$^{\prime\prime}$\\
2020 Oct 8 & 00:05/00:37 &  18 & -1.39 & 1.46$^{\prime\prime}$\\ \hline
\end{tabular}
\end{center}
\end{table}

\section{O$_2$ Mapping} \label{sec:mapping}
Constraining the geography of condensed O$_2$ on Ganymede requires measuring the amount of O$_2$ absorption in each HST spectrum. For the most reliable mapping, we focus on the stronger 5773 \text{\AA} band in our analysis. Initially, we attempt to measure it using individual spectra that had been divided by a standard zero-airmass solar spectrum. However, we find that dividing all of the data by a base spectrum constructed from Ganymede spectra interpreted to contain little O$_2$ provides better cancellation of solar lines and instrument artifacts, allowing for more reliable measurement of the weak ($\leq$ 3$\%$) features.

{We take an iterative approach in constructing an approximately-zero-O$_2$ reference spectrum.} We first construct an initial average spectrum by dividing {all of the spectra by their collective mean}, {dividing out any spectral slopes} across the 5350 -- 6000 \text{\AA} region, measuring the standard deviations of these spectra across the same range, and then including any {pixels} deemed not to contain strong features {(those with a standard deviation $<$ 0.005; 50 spectra total)} in our average. We then divide all of the Ganymede spectra by this {initial} mean and measure the integrated band area of the 5773 \text{\AA} O$_2$ absorption, resulting in both positive and negative measured band strengths. From these measurements, we identify the 5 highest-quality spectra from the 10 leading-hemisphere spectra with the least O$_2$ absorption---{all} below the level of the noise when divided by the solar spectrum---and take the mean as our new, approximately-zero-O$_2$ base spectrum. {The five pixels used in this mean are centered at (104$\degree$W, 32$\degree$N), (117$\degree$W, 33$\degree$N), (121$\degree$W, 35$\degree$N), (133$\degree$W, 11$\degree$N), and (140$\degree$W, 11$\degree$S).} After dividing all of our data by this {final, approximately-zero-O$_2$} average, we re-measure the 5773 \text{\AA} O$_2$ band areas to obtain final values. 

\begin{figure}
\figurenum{1}
\plotone{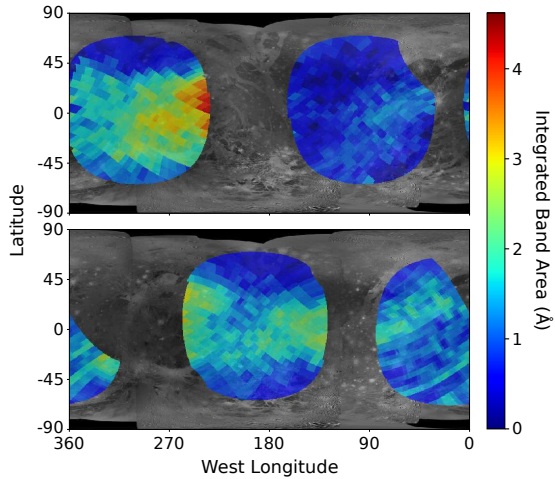}
\caption{Maps of Ganymede's 5773 \text{\AA} condensed O$_2$ absorption for all of the HST/STIS visits. We observe the largest absorptions at the low- to mid-latitudes on the trailing hemisphere. {We separate the observations into two pairs of visits on each panel to facilitate comparison between overlapping observations. The top panel includes the data acquired on 1999 Aug 28 (centered near 300$\degree$W) and 1999 Sep 14 (centered near 97$\degree$W), while the bottom panel includes data acquired on 1999 Sep 23 (centered near 193$\degree$W) and 2020 Oct 8 (centered near 18$\degree$W).} Discrepancies between overlapping regions from {these} different visits are suggestive of moderate temporal variability in the dense-phase O$_2$. \label{fig:map}}
\end{figure}

We make all of our band-strength measurements using the known band shape of Ganymede's 5773 \text{\AA} feature, in order to minimize the effects of noise and residual artifacts on our obtained distribution. We take the high-quality averaged and smoothed ground-based spectrum of Ganymede's O$_2$ from \citet{Spencer1995} (included in their Figure 7) and {remove the spectral slope in the region of the O$_2$ feature}. We then remove the continuum from each HST spectrum by fitting and removing a second-order polynomial from 5400 to 6000 \text{\AA}, excluding the portion corresponding to the feature {($\sim$5530--5820 \text{\AA})}. \deleted{We widen this bound to {5530 \text{\AA}} for the 2020 observation to avoid an interfering instrument artifact that appeared in those data, but that was absent in the older spectra.} Finally, we scale the ground-based absorption to fit the HST data, minimizing the chi-squared across the span of the O$_2$ feature, and integrate the resulting fit to obtain the O$_2$ band area in each HST spectrum. We estimate the obtained pixel-to-pixel uncertainties to be less than 0.5 \text{\AA} of band area on average for the 1999 data, and up to 1 \text{\AA} for the 2020 data (the sub-Jovian observation), due to the more prominent detector effects. As we have already divided each spectrum by a mean with no discernible O$_2$ band above the  noise, we take any slightly negative band strengths as zero absorption. Using the geographic coordinates as obtained in Section \ref{sec:observations}, we then produce a spatially resolved map of Ganymede's 5773 \text{\AA} O$_2$ absorption (Figure \ref{fig:map}).  

\begin{figure}[h!]
\figurenum{2}
\plotone{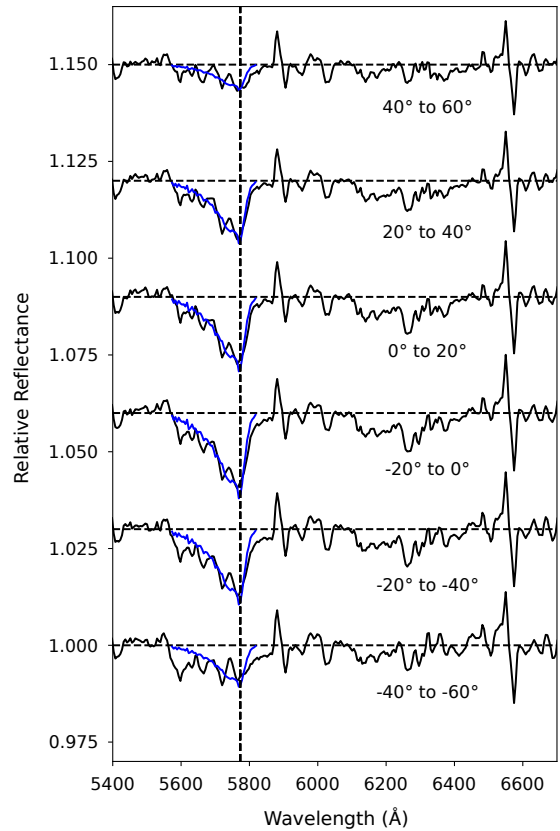}
\caption{Average spectra for different latitude bins across the trailing hemisphere, which demonstrate the changes in strength of the 5773 \text{\AA} feature with latitude. Each spectrum includes all pixels that fall on the trailing hemisphere {(between 180$\degree$W and 360$\degree$W)} and within the designated latitude bin {across all visits}, excluding the few pixels very near the limb of Ganymede. {From top to bottom, the averages include 57, 99, 119, 130, 111, and 97 pixels, respectively.} All spectra have {had polynomial trends removed and been} offset vertically from each other by 0.03 units. The best fits using the feature observed in high-quality ground-based spectra \citep{Spencer1995} are shown overplotted in blue. Horizontal dashed lines indicate the continuum level, and the vertical dashed line indicates the ground-based band minimum at 5773 \text{\AA}. \label{fig:lats}}
\end{figure}

\begin{figure*}[t!]
\figurenum{3}
\plotone{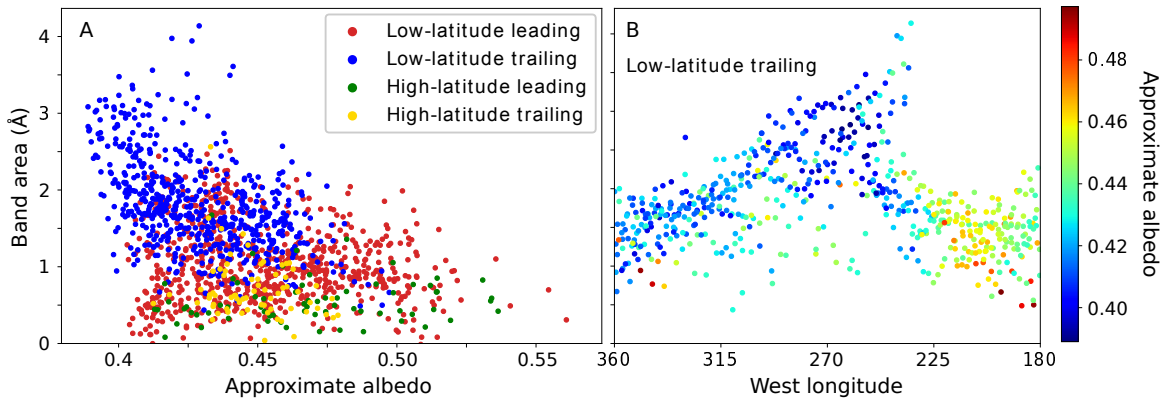}
\caption{Scatter plots investigating the relationships between O$_2$ band strength and albedo (A) and both band strength and albedo with longitude relative to the apex of the trailing hemisphere (B). Pixels are separated by hemisphere and whether they fall equator- or pole-ward of $\pm$40$\degree$ latitude. Only the low-latitude trailing pixels (blue in the left plot) show a slight anti-correlation with albedo. However, this apparent trend is explained by the concurrent slight trend in albedo with longitude from the trailing point to the anti-Jovian point and strong correlation between longitude and band area (B).  \label{fig:albedo}}
\end{figure*}

In agreement with \citet{Spencer1995} and \citet{CalvinSpencer1997}, we find the largest O$_2$ absorptions on the trailing hemisphere and at low latitudes, and relative depletions on the leading hemisphere and pole-ward of roughly $\pm$40$\degree$. Figure \ref{fig:lats} shows example spectra from six latitude bins across the trailing hemisphere (-60$\degree$ to 60$\degree$ in 20$\degree$ increments), demonstrating the variability in band strength with latitude. Aside from this latitudinal trend, which results in weaker O$_2$ features associated with Ganymede's brightened polar caps \citep{Smith1979}, we find no substantial or consistent correlation with albedo. We investigate the variation in the 5773 \text{\AA} band with albedo in Figure \ref{fig:albedo}A, making distinctions between the leading and trailing hemispheres and the low-latitude (equator-ward of $\pm$40$\degree$) and high-latitude (pole-ward of $\pm$40$\degree$) regions. For our purposes, we simply estimate the albedo of each pixel using the USGS \textit{Voyager-Galileo} global mosaic of Ganymede\footnote{https://astrogeology.usgs.gov/search/map/Ganymede/Voyager-Galileo/Ganymede\_Voyager\_GalileoSSI\_global\_mosaic\_1km} and the geometric albedos of the leading and trailing hemispheres \citep{Buratti1991}. We find no correlation within the low-latitude pixels on the leading hemisphere or within those at high latitudes, and only a slight anti-correlation with albedo within the pixels of the low-latitude trailing hemisphere. However, simultaneous inspection of the variation of band strength and albedo with longitude across the trailing hemisphere (Figure \ref{fig:albedo}B) reveals that this trend is readily explained by the coincident increase in band strength and decrease in albedo from the anti-Jovian to the apex of the trailing hemisphere. The apparent lack of correlation between O$_2$ and albedo is again in agreement with the past FOS and WFPC2 observations of \citet{CalvinSpencer1997}. However, contrary to their tentative suggestion that the O$_2$ band may shift to longer wavelengths at higher latitudes, we observe a consistent band minimum at 5773 \text{\AA} within the limits of our data (Figure \ref{fig:lats}). 

\begin{figure}
\figurenum{4}
\plotone{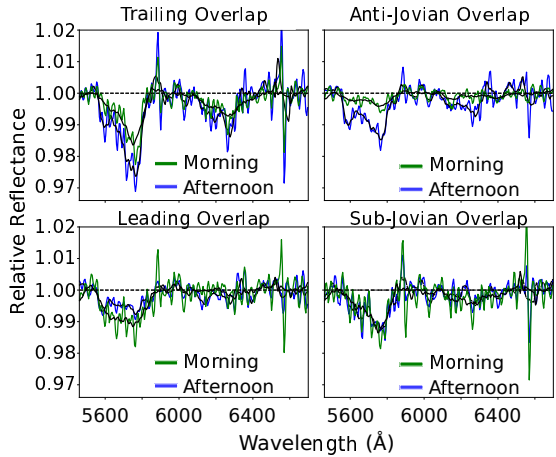}
\caption{Average spectra for regions of overlap between the multiple HST/STIS visits. {The trailing-hemisphere morning and afternoon averages include 48 and 40 pixels, respectively, and span roughly 229--261$\degree$W. The anti-Jovian morning and afternoon averages include 71 and 75 pixels, respectively, and span 123--167$\degree$W. The leading-hemisphere morning and afternoon spectra contain 101 and 104 pixels, respectively, and span 28--87$\degree$W. Finally, the sub-Jovian morning and afternoon overlaps include 20 and 26 pixels, respectively, and span roughly 324--360$\degree$W.} The trailing-hemisphere and anti-Jovian overlaps exhibit larger absorptions in the afternoon than in the morning. However, that trend is reversed in the leading-hemisphere overlap, and no morning/afternoon trend is visible in the sub-Jovian overlap. Thus, while we cannot rule out diurnal fluctuations, we must conclude that the O$_2$ is independently variable due to some other cause. \label{fig:overlaps}}
\end{figure}

Curiously, discrepancies in regions of overlap between different HST/STIS visits appear to show evidence for temporal variability in Ganymede's condensed O$_2$. Overlapping regions on the trailing and anti-Jovian hemispheres from the 1999 data both suggest enhancements in the afternoon relative to the morning, which we initially took for an indication of diurnal variability. However, the new 2020 observation does not show the same trend. Indeed it shows the opposite---stronger absorption in the morning---where it overlaps with the 1999 data on the leading hemisphere, and near perfect agreement on average in the small region of overlap on the sub-Jovian hemisphere. Average spectra of the equatorial ($\pm$30$\degree$) overlapping regions and the entirety of the sub-Jovian overlap, excluding noisier spectra very near the limb, are included in Figure \ref{fig:overlaps}. These averages indicate absolute changes in band depth of up to $\sim$1\%, equivalent to a factor of 3.5 increase for the anti-Jovian overlap, which exhibits the largest relative discrepancy. In addition to these changes appearing robustly above the noise level in average spectra, the large degree of geographic coherence in the apparent variability adds credence to its reality. However, the lack of a consistent diurnal trend suggests that time-of-day fluctuations cannot be the sole explanation for, and may even be unrelated to, the temporal changes we observe.

\section{Discussion} \label{sec:discussion}

The observed equator/pole and leading/trailing dichotomies have implications for the the state of O$_2$ and the mechanisms controlling its production and stability on Ganymede's surface. In agreement with \citet{CalvinSpencer1997} and \citet{Johnson1999comment}, we suggest that the widespread low-latitude enhancement of O$_2$ makes the hypothesis of cold-trapped O$_2$ or of a cold subsurface layer of O$_2$ \citep{VidalEtAl1997, Baragiola1998} unlikely. Instead, trapping mechanisms like the bubble inclusions proposed by \citet{JohnsonJesser1997} seem more consistent. \citet{JohnsonJesser1997} suggest that vacancies formed along the tracks of energetic particles migrate under the surface temperatures relevant to Ganymede to form voids that can trap bubbles of gaseous O$_2$, thereby resulting in the densities necessary to produce the observed O$_2$-dimer absorptions. Their model predicts that O$_2$ production and bubble growth should occur quickest at the warm temperatures of the equatorial latitudes, but that competing sublimation and redeposition effects may actually result in more efficient trapping of O$_2$ at mid-latitudes. Our results showing widespread O$_2$ at low latitudes, but no extra enhancements at the mid-latitudes, are in partial agreement with these predictions.

Overall, the geographic distribution we obtain is suggestive of formation driven by high-energy ions and destruction driven by sputtering from lower-energy ions. Ganymede's intrinsic magnetic field \citep{Kivelson1996} is expected to deflect the vast majority of incident Jovian magnetospheric ions to the high latitudes (pole-ward of approximately $\pm$30--40$\degree$) \citep{Poppe2018}, resulting in the most intense radiation and largest sputtered fluxes in these regions---precisely where we see the least O$_2$ absorption. The low latitudes of the leading hemisphere, which are similarly depleted in O$_2$, are still predicted to receive significantly more radiation and experience higher sputtering rates than the equatorial latitudes of the trailing hemisphere---the most sheltered region of Ganymede's surface and where we see the largest O$_2$ bands. Only high-energy, penetrating ions are thought to access the equatorial latitudes of the trailing hemisphere \citep{Poppe2018} to provide the necessary energy for radiolysis and O$_2$ formation. These patterns are again consistent with the hypothesis of \citep{JohnsonJesser1997}, who suggest that energetic heavy ions may efficiently produce O$_2$ and that sputtering may effectively destroy the bubble inclusions.  

It is interesting to compare the geography of Ganymede's O$_2$ to that of its ozone (O$_3$), which results from similar radiolytic processes and causes an absorption feature near 2600 \text{\AA}. Like O$_2$, the O$_3$ appears more strongly on the trailing hemisphere \citep{Noll1996}. However, limited spatially resolved observations from the \textit{Galileo} Ultraviolet Spectrometer show enhancements at the poles and morning and evening limbs, which have been interpreted to reflect the conversion of O$_3$ to O$_2$ by UV photolysis at low solar zenith angles \citep{Hendrix1999}. It is not clear how or whether the apparent lack of O$_3$ at midday low-latitudes relates to the O$_2$ we find there, as we do not see any consistent trend with solar zenith angle. However, the potential temporal variability in O$_2$ that we observe suggests that dynamic processes may also be affecting this species.

As noted in Section \ref{sec:mapping}, we initially interpreted afternoon enhancements of the O$_2$ bands in overlapping regions of the 1999 data as evidence for potential diurnal variability. If diurnal fluctuations were indeed acting to enhance O$_2$ concentrations in the afternoon, that could be consistent with bubble aggregation and growth under warmer temperatures as laid out by \citet{JohnsonJesser1997}. However, the addition of the 2020 visit, which shows the opposite morning/afternoon trend on the leading hemisphere and no trend in the small overlapping portion of the sub-Jovian hemisphere, indicates that there must be an alternate source of variability. The observations presented here constrain the timescale of this variability to be $\leq$9 days---the shortest gap between overlapping observations. With the available data, we cannot distinguish between more stochastic causes, such as temporal fluctuations in the incident particle flux or composition, and a combination of such a cause and potential diurnal effects.

{Though the overlapping edges of each observation were viewed at similar geometries, we also cannot entirely exclude the possibility of preferential O$_2$ concentration on west-facing slopes, which could, in principle, contribute to the enhanced band strengths seen on the eastern (afternoon) sides of the trailing and anti-Jovian visits. Differences in surface color between west- and east-facing slopes have been seen on Saturn's moons Dione and Tethys \citep{Schenk2011}, for example, but it is unclear how large of an effect this would produce on Ganymede or why the leading overlap would then show the opposite trend.} Further observations of Ganymede's condensed O$_2$ may help confirm the variability and distinguish between {geometric, diurnal, and stochastic} effects. 

\section{Conclusions} \label{sec:conclusions}
Using HST/STIS spectroscopy, we have mapped
Ganymede’s 5773 \text{\AA} dense-phase O$_2$ absorption across its surface. In agreement with past observations, we find the strongest absorptions at low- to mid-latitudes on the trailing hemisphere and weaker absorptions at the high latitudes and on the leading hemisphere. We interpret this distribution to reflect the influence of Ganymede's intrinsic magnetic field in sheltering the trailing-hemisphere equatorial regions from charged particle sputtering. In addition to mapping its geography, we also observe potential temporal variability in Ganymede's surface O$_2$ in the form of band-strength discrepancies in regions of overlap between the HST visits. We suggest temporal variation in the local magnetospheric environment as a possible cause, but we cannot distinguish between potential contributing factors with only the data at hand.

\acknowledgments
Based on observations made with the NASA/ESA Hubble Space Telescope, obtained from the Data Archive at the Space Telescope Science Institute, which is operated by the Association of Universities for Research in Astronomy, Inc., under NASA contract NAS5-26555. These observations are associated with program $\#$7444 and $\#$15789. This work is also based on new observations associated with program $\#$15925. Support for program $\#$15925 and $\#$15789 was provided by NASA through a grant from the Space Telescope Science Institute, which is operated by the Association of Universities for Research in Astronomy, Inc., under NASA contract NAS5-26555. This work was also supported by NASA Headquarters under the NASA Earth and Space Science Fellowship Program (grant 80NSSC17K0478). {We would like to acknowledge John R. Spencer for his role in producing the 1999 observations used in this study.}


\software{astropy \citep{2013A&A...558A..33A}}




\end{document}